\documentclass[twocolumn,showpacs,showkeys,preprintnumbers,amsmath,amssymb]{revtex4}
\usepackage{graphicx}
\usepackage{dcolumn}
\usepackage{bm}

\begin{document}

\preprint{APS/123-QED}

\title{Breaking a secure communication scheme based
on the phase synchronization of chaotic systems}
\author{G. \'{A}lvarez}
\email{gonzalo@iec.csic.es}
\author{F. Montoya}
\author{G. Pastor}
\author{M. Romera}
\affiliation{
Instituto F\'{\i}sica Aplicada\\
Consejo Superior de Investigaciones Cient\'{\i}ficas\\
Serrano 144, 28006--Madrid, Spain}

\date{\today}


\begin{abstract}
A security analysis of a recently proposed secure communication
scheme based on the phase synchronization of chaotic systems is
presented. It is shown that the system parameters directly
determine the ciphertext waveform, hence it can be readily broken
by parameter estimation of the ciphertext signal.

\end{abstract}

\maketitle

Most secure chaotic communication systems are based on complete
synchronization (CS), whereas a new cryptosystem has been proposed
based on phase synchronization (PS). This scheme hides binary
messages in the instantaneous phase of the drive subsystem used as
the transmitting signal to drive the response subsystem. Although
it is claimed to be secure against some traditional attacks in the
chaotic cryptosystems literature, including the parameter
estimation attack, we show that it is breakable by this attack. As
a conclusion, the system is not secure and should not be used for
communications where security is a strict requirement.

\section{Introduction}

In recent years, a great number of cryptosystems based on chaos
have been proposed \cite{asocscs,cc}, most of them fundamentally
flawed by a lack of robustness and security
\cite{stusc,cocborcr,emmbc,uamccs,pwtcisea,bcsugse,bcscuas,ccscurm,coaces,coacscs,coaecc,otsoacespwccifcp}.
In~\cite{ascsbotpsocs}, a secure communication scheme based on the
phase synchronization of a chaotic system is proposed.

In this new scheme the plaintext binary message $b$ is hidden in
the instantaneous phase of the drive subsystem used as
transmitting signal to drive the response subsystem. At the
response subsystem, the phase difference is detected and its
strong fluctuation above or below zero recovers the plaintext at
certain coupling strength.

The secure communication process is illustrated by means of an
example based on coupled R\"ossler chaotic oscillators. In the
example, the drive subsystem is formed by two weak coupled
oscillators. The plaintext is used to modulate the same parameter
in both oscillators 1 and 2. The equations of the drive subsystem
are:

\begin{eqnarray}
&&\dot{x}_{1,2}=-(\omega+\Delta\omega)y_{1,2}-z_{1,2}+
\varepsilon(x_{2,1}-x_{1,2}),\nonumber\\
&&\dot{y}_{1,2}=(\omega+\Delta\omega)x_{1,2}+\alpha y_{1,2},\\
&&\dot{z}_{1,2}=\beta+z_{1,2}(x_{1,2}-\gamma).\nonumber
\end{eqnarray}

The response subsystem is governed by:
\begin{eqnarray}
&&\dot{x}_{3}=-\omega\,'y_{3}-z_{3}+
\eta((x^2_3+y_3^2)^{1/2}cos\phi_m-x_3),\nonumber\\
&&\dot{y}_{3}=\omega\,'x_{3}+\alpha\,' y_{3},\\
&&\dot{z}_{3}=\beta+z_{3}(x_{3}-\gamma).\nonumber
\end{eqnarray}

\begin{figure}[t]
\begin{center}
  \includegraphics{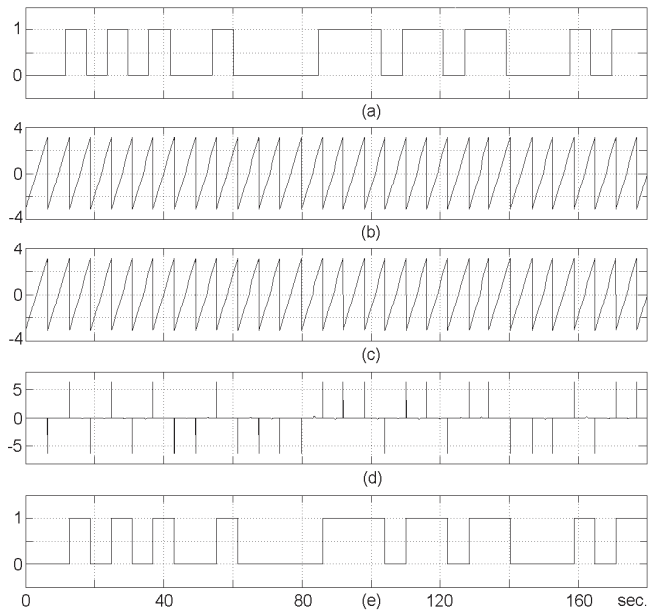}
  \caption{Plaintext recovery with the authorized receiver. Time
  histories of: (a) plaintext $b$; (b) ciphertext $\phi_m^*$;
  (c) reconstructed phase signal of the response subsystem
  $\phi_3^*$; (e) difference between the ciphertext and the
  reconstructed signal $\phi_m^*-\phi_3^*$; (f) reconstructed
  plaintext $b'$.} \label{fig:legal}
\end{center}
\end{figure}

In the example, the parameter values are: $\omega=\omega\,'=1,$
$\varepsilon=5\times10^{-3}$, $\eta=5.3$, and
$\alpha=\alpha\,'=0.15\,.$

The parameters $\beta$ and $\gamma$ are held as constants, with
the values $\{\beta,\gamma\}=\{0.2, 10\}$.

The parameter $\omega$ corresponds to the natural frequency of the
R\"ossler oscillator drive subsystems 1 and 2. The parameter
$\omega\,'$ corresponds to the natural frequency of the R\"ossler
oscillator driven subsystem 3, $\varepsilon$ corresponds to the
weak coupling factor between the oscillators 1 and 2, and $\eta$
corresponds to the strong coupling factor in the driven oscillator
3.

The parameter mismatch $\Delta\omega$ is modulated by the
plaintext, being $\Delta\omega=0.01$ if the bit to be transmitted
is ``1'' and $\Delta\omega=-0.01$ if the bit to be transmitted is
``0''.

\begin{figure}[t]
\begin{center}
  \includegraphics{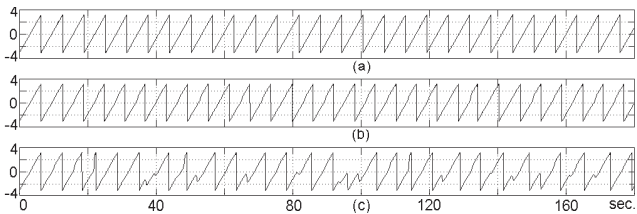}
  \caption{Ciphertext phase signal $\phi_m^*$ as function of $\alpha$: (a)
  $\alpha=0.01$, the phase increases almost linearly; (b)
 $\alpha=0.15$, the phase increases monotonically with chaotic
 behavior; (c) $\alpha=0.25$, the phase increases and decreases
 irregularly.}
  \label{fig:alfa}
\end{center}
\end{figure}

The ciphertext consists of the phase of the mean field of the
drive oscillators:

\[\phi_m=\arctan\frac{x_1+x_2}{y_1+y_2}.\]

As the phase is a signal that has an unbounded amplitude it can
not be transmitted through physical channels. This problem is
overcome by coding the signal from $\pi$ to $-\pi$, which
corresponds to the Poincar\'{e} surface of the atractor,
$y_{1,2}=0$. As a consequence, the transmitted ciphertext, marked
as $\phi_m^*$, is a sawtooth-like signal with a period equal to
the revolution period of the oscillator.

At the receiving end the phase of the response subsystem is:

\[\phi_3=\arctan\frac{x_3}{y_3}\]

that is also coded from $\pi$ to $-\pi$ as $\phi_3^*$.

The plaintext is retrieved by calculating the difference between
the ciphertext and the reconstructed signal, $\phi_m^*-\phi_3^*.$
The difference signal consisted of positive and negative peaks
that correspond to the ones and zeros of the plaintext.

The example of \cite{ascsbotpsocs} is illustrated in
Fig.~\ref{fig:legal}. We have simulated it with a four order
Runge-Kutta integration algorithm in MATLAB 6, with a step size of
$0.001$. In order to recover the plaintext with the exact
waveform, allowing for a small time delay, we have included a
Smith-trigger as a reconstruction filter, with switch on point at
4 and switch off point at -4.

As in the example of \cite{ascsbotpsocs} there is no indication
about the parameter initial values, our simulation is implemented
with the following initial values: $(x_1^{(0)}, x_2^{(0)},
x_3^{(0)}, y_1^{(0)}, y_2^{(0)}, y_3^{(0)}, z_1^{(0)}, z_2^{(0)},
z_3^{(0)})=(-5, -3, -1, 0, 0, 0, 0, 0, 0)$.

The authors seemed to base the security of its secure
communication system on the properties of the phase
synchronization. They claimed that it can not be broken by some
traditional attacks used against secure chaotic systems with
complete synchronization, but no general analysis of security was
included.

Although the authors point out that the system parameters play the
role of secret key in transmission~\cite[ \S V]{ascsbotpsocs}, it
is not clearly specified which parameters are considered as
candidates to form part of the key, what the allowable value range
of those parameters is, what the key space is (how many different
keys exist in the system) and how they would be managed.

The weaknesses of this system and the method to break it are
discussed in the next section.

\begin{figure}[t]
  \center
  \includegraphics{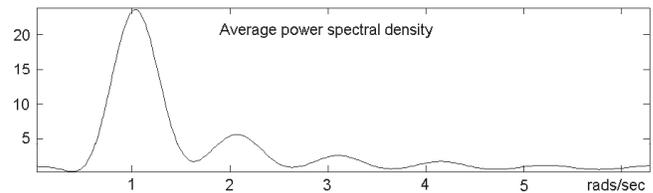}
  \caption{Power spectral analysis of the ciphertext
  signal. The highest peak corresponds to the frequency of $\omega$
  and lies at $\omega\approx 1.1\,.$}
  \label{fig:espectro}
\end{figure}

\section{Breaking the system}
\label{sec:breaking}

The main problem with this cryptosystem lies on the fact that the
ciphertext is an analog signal, whose waveform depends on the
system parameter values. Likewise, the difference between the
ciphertext and the phase signal of a non synchronized receiver
$\phi_m^*-\phi_3^*$, depends on these same parameters. The study
of these signals provides the necessary information to recover a
good estimation of the system parameter values and the correct
plaintext, as will be seen next.

Let us assume that the key consists of the oscillator's parameters
$\alpha$ and $\omega$, as they are the only unknowns in the
example of \cite{ascsbotpsocs}. Moreover the parameters $\beta$
and $\gamma$, that were constants in the example, can not be part
of the key because, according to our experiments, the
synchronization of the R\"ossler oscillator is indifferent to a
mismatch of the value of these parameters in a range greater than
1 to 1000.

The search space of $\alpha$ may be restricted to the unique
suitable value range for operation, characterized by the mild
chaotic region of the R\"ossler oscillator, in which its phase
increases monotonically with time, showing a chaotic increase
rate, that allows hiding the binary information. This region is
roughly characterized by the following values of $\alpha$:

\begin{equation}\label{eq:chaotic}
0.03\leq\alpha\leq 0.18\;.
\end{equation}

The operation of the system with lower values of $\alpha$ should
be avoided because the waveform of the oscillator is quite uniform
and its phase increases almost linearly with time. Therefore, the
instantaneous phase fluctuations, due to the binary information
modulation, can not be effectively hidden, and thus the
information could be easily retrieved from the signal.

Higher values of $\alpha$ should  be also avoided because the
R\"ossler oscillator operates in the wild chaotic region, in which
the phase does not increases monotonically with time, showing
erratic increases and decreases, rendering impossible the
synchronization of the authorized receiver, thus preventing the
correct data retrieving.

\begin{figure}[t]
\begin{center}
  \includegraphics{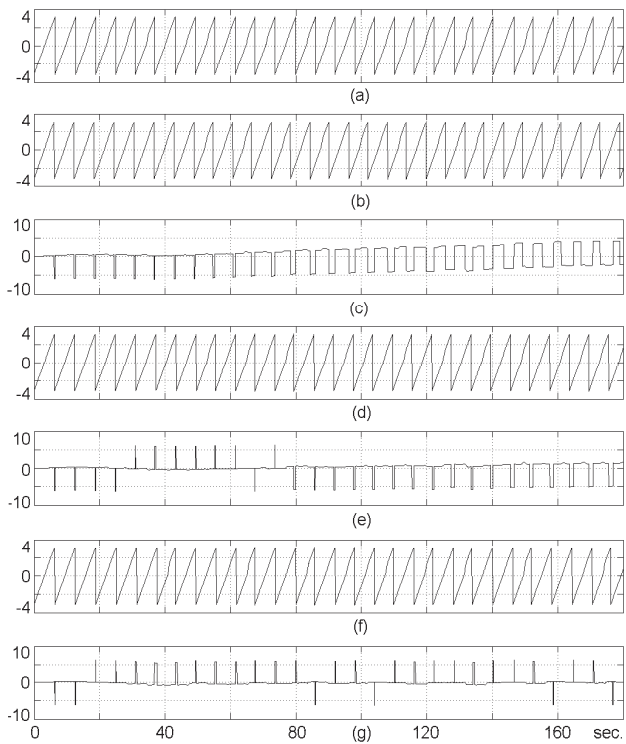}
  \caption{Determination of the best value of ${\omega}'$: (a)
ciphertext signal with frequency $\omega=1.00$; (b) phase signal
of the free running intruder receiver $\phi_3^*$ for
${\omega}'=1.03$; (c) output of the phase comparator $\phi_m^* -
\phi_3^*$ for ${\omega}'=1.03$; (d) phase signal of the free
running intruder receiver $\phi_3^*$ for ${\omega}'=1.015$; (e)
output of the phase comparator $\phi_m^* - \phi_3^*$ for
$\omega'=1.015$; (f) phase signal of the free running intruder
receiver $\phi_3^*$ for $\omega'=1.005$; (g) output of the phase
comparator $\phi_m^* - \phi_3^*$ for $\omega'=1.005$.}
  \label{fig:omega}
\end{center}
\end{figure}

The behavior of the attractor with respect to $\alpha$ is
illustrated in Fig.~\ref{fig:alfa}, in which the time history of
the ciphertext signal $\phi_m^*$ for three values of $\alpha$ is
shown. The first sample corresponds to $\alpha=0.01$, showing that
the phase increases almost linearly. The second one corresponds to
$\alpha=0.15$, showing that the phase increases monotonically with
chaotic behavior. The last sample corresponds to $\alpha=0.25$,
showing that the phase increases and decreases irregularly.

The sensitivity to the parameter values is so low that the
original plaintext can be recovered from the ciphertext using an
intruder receiver system with parameter values considerably
different from the ones used by the transmitter (\cite[Fig.
7]{ascsbotpsocs}).

We have found that the plaintext $b\,'$ can be recovered even when
$\alpha\,'$ has an absolute error of $\pm\, 0.2$. As a
consequence, it is sufficient to try four values of $\alpha\,'$,
to cover its full usable range. The best set of values is:
$\alpha\,'=\{0.05, 0.09, 0.13, 0.17\}$.

In Fig.~\ref{fig:espectro} we show the power spectral analysis of
the ciphertext signal. As can be observed, the frequency of the
R\"ossler oscillator is totally evident. The spectrum's highest
peak appears at $\omega\,'\simeq 1.03$, close to the parameter
value of the drive subsystem $\omega=1$. Thus, by simply examining
the ciphertext, the second key element $\omega\,'$ is guessed with
reasonable accuracy.

Let ${\omega}\,'$ be the approximate value of $\omega$. Once it is
measured we can use it to recover the plaintext in the following
way.

\begin{figure}[t]
\begin{center}
  \includegraphics{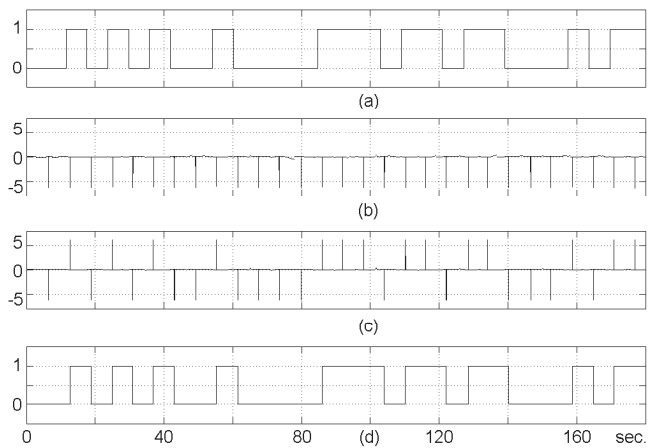}
  \caption{Determination of the best value of $\alpha\,'$, with
$\omega\,'=1.005$: (a) original plaintext, $b$; (b) output of the
phase comparator $\phi_m^* - \phi_3^*$ for $\alpha\,'=\{0.05,
0.09, 0.13\}$, which is the same in three cases; (c) output of the
phase comparator $\phi_m^* - \phi_3^*$ for $\alpha\,'=0.17$; (d)
recovered plaintext $b'$ for $\alpha\,'=0.17$.}
\label{fig:recupera}
\end{center}
\end{figure}

First, we introduce the estimated value of ${\omega}\,'$ into an
intruder receiver with $\eta=0$, that is without coupling, so the
intruder receiver oscillator will be running freely. To check
whether the estimation of ${\omega}\,'$ is good, we look at the
output of the phase comparator $\phi_m^* - \phi_3^*$ as well as at
the ciphertext signal $\phi_m^*$ and at the phase signal of the
receiver $\phi_3^*$.

When the frequencies of transmitter and intruder receiver are
slightly different, then $\phi_m^* - \phi_3^*$ will look like a
train of pulses of increasing width summed with a direct current
of increasing level; being the final width and direct current
increasing level rate proportional to the difference of
frequencies $\omega\,'-\omega$. Also, the mismatch of the periods
of the phase signals $\phi_m^*$ and $\phi_3^*$ is perceptible.
With this information we can adjust the value of $\omega\,'$ in a
few steps, until the width of the pulses tends to zero. Then, the
period mismatch of the phase signals $\phi_m^*$ and $\phi_3^*$ is
unnoticeable and its direct current level equals zero.

The procedure is illustrated in Fig.~\ref{fig:omega}. We begin
with $\omega\,'=1.03$, the value estimated from the spectrum, and
we see that the correct value of $\omega\,'$ must be slightly
lower, thus we try $\omega\,'=1.015$ and we see that we are near
the exact value but still a little bit high. Next we try
$\omega\,'=1.005$, and we see that the frequency match is quite
good. We retain this last value of ${\omega}\,'$ as the definite
one and go to the next step.

Finally, we set $\eta=5.3$ at the intruder receiver and look at
the retrieved data $b\,'$ for the previously obtained
${\omega}\,'$ an for each of the four possible values of
$\alpha\,'$. In Fig.~\ref{fig:recupera} the retrieved binary data
$b\,'$ obtained with $\omega\,'=1.005$ and $\alpha\,'=\{0.05,
0.09, 0.13, 0.17\}$ are presented. It can be seen that for
$\alpha\,'=\{0.05, 0.09, 0.13\}$ only cero value data are obtained
and for $\alpha\,'=0.17$ some output data are present, thus we may
assume that the value of $\alpha\,'= 0.17$ can be retained as the
appropriate one to retrieve the plaintext $b\,'$ and that the data
obtained with it consist of the correct recovered plaintext, as
can be verified from the figure.

\begin{figure}[t]
\begin{center}
  \includegraphics{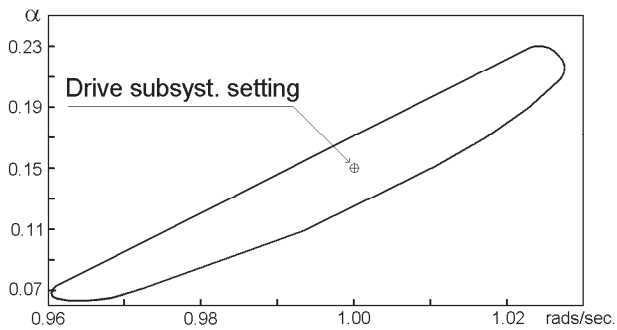}
  \caption{Range of $\omega\,'$ and $\alpha\,'$ values that that
   achieve correct plaintext recovery of a ciphertext generated
   with $\{\omega, \alpha\}=\{1.00, 0.15\}$.}
   \label{fig:enganche}
\end{center}
\end{figure}

Although the estimated pair of values $\{\omega\,',
\alpha\,'\}=\{1.005, 0.17\}$ are far from the right ones, the
plaintext is correctly recovered as a consequence of the system's
low sensitivity to parameters.

Moreover, we have observed that many other combinations of
parameter values allow for the recovery of the correct plaintext
as well. In Fig.~\ref{fig:enganche} we show after many simulations
the region of $\{\omega\,', \alpha\,'\}$ values in which correct
plaintext recovery of a ciphertext generated with a drive
subsystem with $\{\omega, \alpha\}=\{1.00, 0.15\}$ is achieved.

\section{Conclusion}
The proposed cryptosystem is rather weak, since it can be broken
by measuring the power spectrum of the ciphertext signal and
trying a small set of parameter values. There is no detailed
description about what the key is, nor what the key space is, a
fundamental aspect in every secure communication system. The lack
of security discourages the use of this algorithm for secure
applications.

\vspace{0.5cm} \noindent {\bf Acknowledgements} This work is
supported by \textit{Ministerio de Ciencia y Tecnolog\'{\i}a} of
Spain, research grant TIC2001-0586.

\end{document}